\documentclass[12pt]{cernart}
\usepackage{epsfig}

\newcommand{\beq}  {\begin{equation}}
\newcommand{\eeq}  {\end{equation}}
\newcommand{\bmath}{\begin{eqnarray}}
\newcommand{\emath}{\end{eqnarray}}
\newcommand{\bei}{\begin{itemize}}
\newcommand{\eei}{\end{itemize}}
\newcommand{\ks}{K_S}
\newcommand{\kl}{K_L}
\newcommand{\gaga}{\gamma\gamma}
\newcommand{\kgg}{K~\rightarrow~\gamma\gamma}
\newcommand{\ksgg}{K_{S}\rightarrow\gamma\gamma}
\newcommand{\klgg}{K_{L}\rightarrow\gamma\gamma}
\newcommand{\kspp}{K_{S}\rightarrow{\pi^0}{\pi^0}}
\newcommand{\klppp}{K_{L}\rightarrow{\pi^0}{\pi^0}{\pi^0}}
\newcommand{\pipipin}{{\pi^0}{\pi^0}{\pi^0}}
\newcommand{\pipin}{{\pi^0}{\pi^0}}
\newcommand{\ggtoppp}{\frac{\Gamma(\klgg)}{\Gamma(\klppp)}}

\begin{document}
\begin{titlepage}
\docnum{CERN--EP/2002-074}
\date{10. 10. 2002}
\title{\bf \Large Precise measurements of the $\ksgg$ and~$\klgg$ decay rates}
\begin{Authlist}
\begin{center}
{\bf NA48 collaboration}
\  \\[0.2cm] 
 A.~Lai,
 D.~Marras \\
{\em \small Dipartimento di Fisica dell'Universit\`a e Sezione dell'INFN di Cagliari, I-09100 Cagliari, Italy} \\[0.2cm] 
J.R.~Batley,
 R.S.~Dosanjh,
 T.J.~Gershon,
G.E.~Kalmus,
 C.~Lazzeroni,
 D.J.~Munday,
E.~Olaiya,
 M.A.~Parker,
 T.O.~White,
 S.A.~Wotton \\
{\em \small Cavendish Laboratory, University of Cambridge, Cambridge, CB3 0HE, U.K.\footnotemark[1]} \\[0.2cm] 
R.~Arcidiacono\footnotemark[2],
G.~Barr,
 G.~Bocquet,
 A.~Ceccucci,
 T.~Cuhadar-D\"onszelmann,
 D.~Cundy\footnotemark[3],
 N.~Doble,
V.~Falaleev,
 L.~Gatignon,
 A.~Gonidec,
 B.~Gorini,
 P.~Grafstr\"om,
W.~Kubischta,
 A.~Lacourt,
 I.~Mikulec\footnotemark[4],
A.~Norton,
 B.~Panzer-Steindel,
G.~Tatishvili\footnotemark[5],
 H.~Wahl \\
{\em \small CERN, CH-1211 Gen\`eve 23, Switzerland} \\[0.2cm] 
C.~Cheshkov,
 P.~Hristov,
V.~Kekelidze,
 D.~Madigojine,
N.~Molokanova,
Yu.~Potrebenikov,
 A.~Zinchenko \\
{\em \small Joint Institute for Nuclear Research, Dubna, Russian    Federation} \\[0.2cm] 
%
%
 V.~Martin,
 P.~Rubin,
 R.~Sacco,
 A.~Walker \\
{\em \small Department of Physics and Astronomy, University of    Edinburgh, JCMB King's Buildings, Mayfield Road, Edinburgh,    EH9 3JZ, U.K.} \\[0.2cm] 
%
%
M.~Contalbrigo,
 P.~Dalpiaz,
 J.~Duclos,
P.L.~Frabetti,
 A.~Gianoli,
 M.~Martini,
 F.~Petrucci,
 M.~Savri\'e \\
{\em \small Dipartimento di Fisica dell'Universit\`a e Sezione    dell'INFN di Ferrara, I-44100 Ferrara, Italy} \\[0.2cm] 
%
%
A.~Bizzeti\footnotemark[6],
M.~Calvetti,
 G.~Collazuol,
 G.~Graziani,
 E.~Iacopini,
 M.~Lenti,
 F.~Martelli,
 M.~Veltri \\
{\em \small Dipartimento di Fisica dell'Universit\`a e Sezione    dell'INFN di Firenze, I-50125 Firenze, Italy} \\[0.2cm] 
%
%
 M.~Eppard,
 A.~Hirstius,
 K.~Holtz,
 K.~Kleinknecht,
 U.~Koch,
 L.~K\"opke,
 P.~Lopes da Silva, 
P.~Marouelli,
 I.~Mestvirishvili,
 I.~Pellmann,
 A.~Peters,
S.A.~Schmidt,
  V.~Sch\"onharting,
 Y.~Schu\'e,
 R.~Wanke,
 A.~Winhart,
 M.~Wittgen \\
{\em \small Institut f\"ur Physik, Universit\"at Mainz, D-55099    Mainz, Germany\footnotemark[7]} \\[0.2cm] 
J.C.~Chollet,
 L.~Fayard,
 L.~Iconomidou-Fayard,
 G.~Unal,
 I.~Wingerter-Seez \\
{\em \small Laboratoire de l'Acc\'el\'erateur Lin\'eaire,  IN2P3-CNRS,Universit\'e de Paris-Sud, 91898 Orsay, France\footnotemark[8]} \\[0.2cm] 
G.~Anzivino,
 P.~Cenci,
 E.~Imbergamo,
 P.~Lubrano,
 A.~Mestvirishvili,
 A.~Nappi,
M.~Pepe,
 M.~Piccini \\
{\em \small Dipartimento di Fisica dell'Universit\`a e Sezione    dell'INFN di Perugia, I-06100 Perugia, Italy} \\[0.2cm] 
%
%
R.~Casali,
 C.~Cerri,
 M.~Cirilli\footnotemark[9],
F.~Costantini,
 R.~Fantechi,
 L.~Fiorini,
 S.~Giudici,
 I.~Mannelli,
G.~Pierazzini,
 M.~Sozzi \\
{\em \small Dipartimento di Fisica, Scuola Normale Superiore e Sezione dell'INFN di Pisa, I-56100 Pisa, Italy} \\[0.2cm] 
%
%
J.B.~Cheze,
 M.~De Beer,
 P.~Debu,
 F.~Derue,
 A.~Formica,
 R.~Granier de Cassagnac,
G.~Gouge,
G.~Marel,
E.~Mazzucato,
 B.~Peyaud,
 R.~Turlay,
 B.~Vallage \\
{\em \small DSM/DAPNIA - CEA Saclay, F-91191 Gif-sur-Yvette, France} \\[0.2cm] 
M.~Holder,
 A.~Maier,
 M.~Ziolkowski \\
{\em \small Fachbereich Physik, Universit\"at Siegen, D-57068 Siegen, Germany\footnotemark[10]} \\[0.2cm] 
 C.~Biino,
 N.~Cartiglia,
 F.~Marchetto, 
E.~Menichetti,
 N.~Pastrone \\
{\em \small Dipartimento di Fisica Sperimentale dell'Universit\`a e    Sezione dell'INFN di Torino,  I-10125 Torino, Italy} \\[0.2cm] 
J.~Nassalski,
 E.~Rondio,
 M.~Szleper,
 W.~Wislicki,
 S.~Wronka \\
{\em \small Soltan Institute for Nuclear Studies, Laboratory for High    Energy Physics,  PL-00-681 Warsaw, Poland\footnotemark[11]} \\[0.2cm] 
H.~Dibon,
 M.~Jeitler,
 M.~Markytan,
 G.~Neuhofer,
M.~Pernicka,
 A.~Taurok,
 L.~Widhalm \\
{\em \small \"Osterreichische Akademie der Wissenschaften, Institut  f\"ur Hochenergiephysik,  A-1050 Wien, Austria\footnotemark[12]} \\[1cm] 
\end{center}
\end{Authlist}

\begin{abstract}
The $\ksgg$ decay rate has been measured with the NA48 detector
using a high intensity short neutral beam from the CERN SPS. The measured
branching ratio \mbox{$BR(\ksgg) = (2.78 \pm 0.06_{stat} \pm 0.04_{syst})
\times 10^{-6}$}, obtained from 7461$\pm$172 $\ksgg$ events, is
significantly higher than the $O(p^4)$ prediction of Chiral Perturbation
Theory. Using a $\kl$ beam the ratio $\ggtoppp = (2.81
\pm 0.01_{stat} \pm 0.02_{syst}) \times 10^{-3}$ has been measured.
\end{abstract}

\maketitle

\footnotetext[1]{ Funded by the U.K.    Particle Physics and Astronomy Research Council}
\footnotetext[2]{ On leave from Dipartimento di Fisica Sperimentale 
dell'Universit\`a e    Sezione dell'INFN di Torino,  I-10125 Torino, Italy}
\footnotetext[3]{Present address: Instituto di Cosmogeofisica del CNR di Torino, I-10133 Torino, Italy}
\footnotetext[4]{ On leave from \"Osterreichische Akademie der Wissenschaften, Institut  f\"ur Hochenergiephysik,  A-1050 Wien, Austria}
\footnotetext[5]{ On leave from Joint Institute for Nuclear Research, Dubna,141980, Russian Federation}
\footnotetext[6]{ Dipartimento di Fisica dell'Universita' di Modena e Reggio Emilia, via G. Campi 213/A I-41100, Modena, Italy}
\footnotetext[7]{ Funded by the German Federal Minister for    Research and Technology (BMBF) under contract 7MZ18P(4)-TP2}
\footnotetext[8]{ Funded by Institut National de Physique des
 Particules et de Physique Nucl\'eaire (IN2P3), France}
\footnotetext[9]{Present address: Dipartimento di Fisica
 dell'Universit\'a di Roma ``La Sapienza'' e Sezione dell'INFN di Roma,
 00185 Roma, Italy}
\footnotetext[10]{ Funded by the German Federal Minister for Research and Technology (BMBF) under contract 056SI74}
\footnotetext[11]{Supported by the Committee for Scientific Research grants
5P03B10120, SPUB-M/CERN/P03/DZ210/2000 and SPB/CERN/P03/DZ146/2002}
\footnotetext[12]{    Funded by the Austrian Ministry for Traffic and Research under the    contract GZ 616.360/2-IV GZ 616.363/2-VIII, and by the Fonds f\"ur   Wissenschaft und Forschung FWF Nr.~P08929-PHY}

\end{titlepage}

\setcounter{footnote}{0}

\section{\bf Introduction}
The decays $K_{S,L}\rightarrow\gamma\gamma$ and 
$K_{S,L}\rightarrow{\pi^0}\gamma\gamma$ are important probes 
of Chiral Perturbation Theory (${\chi}PT$), an effective field theory
of the Standard Model at low energies. The decay amplitude of $\ksgg$ 
can be calculated unambiguously at the leading $O(p^4)$
order of the perturbative expansion giving a branching ratio of $2.1
\times 10^{-6}$ with an uncertainty of only a few per
cent~\cite{theor}\footnote{These results are similar to those obtained
  in earlier calculations based on a phenomenological approach~\cite{thold}}. 
The theoretical predictions for  $\klgg$ are much less accurate~\cite{kltheo}.
The $\ksgg$ decay rate has hitherto been
considered as an example of good agreement between
the ${\chi}PT$ prediction and the experimental measurement
$(2.4 \pm 0.9) \times 10^{-6}$, obtained by the NA31 experiment~\cite{na31}. 
This result was recently improved by 
NA48 to $(2.58 \pm 0.42) \times 10^{-6}$ \cite{na48} with data from a two-day
test run with high intensity short neutral beam, collected in September 1999. 
However, the previous experimental results are not precise enough
to resolve the higher order effects which are
predicted by the ${\chi}PT$ to be at most of the order of
$m_K^2/(4\pi F_{\pi})^2 \sim 20\%$ of the $O(p^4)$ decay
amplitude~\cite{theor}. Using data taken in a 40 day run with the high 
intensity short neutral beam in 
the year 2000, for the first time an accuracy of 
a few~\% has been achieved. A comparison of the experimental result with
the $O(p^4)$ ${\chi}PT$ prediction allows the estimation of the
$O(p^6)$ loop contributions. According to ${\chi}PT$, these could
also contribute to $O(p^6)$ effects in the 
$K_{L}\rightarrow{\pi^0}\gamma\gamma$ decay which up to now are believed to be
dominated by vector meson exchange.

\section{Principle of the measurement}
The NA48 experiment has been designed to measure direct CP violation 
in neutral kaon decays and comprises $\ks$ and $\kl$ beams
created at two targets, near and far from the decay region, which
can be run together or separately~\cite{doble}. 

In previous measurements of $\ksgg$,
the decay region used contained almost twice as many decays coming from
$\klgg$. Their subtraction using the $\kl$ flux estimated from e.g. 
$\klppp$ decays introduces an uncertainty of more than 4\% due to
the poorly known $\klgg$ branching ratio.
In the measurement presented here, this problem is overcome by using a
novel method based on the observed ratio of $\klgg$ to $\klppp$, as a
function of energy and decay position, in the $\kl$ beam (far-target) 
run performed immediately 
prior to the $\ks$ beam (near-target) data taking. 
As the near- and far-target beams are almost
collinear and have similar momentum spectra, the measurement of
$\klppp$ decays in  the near-target beam, as a function of energy and decay
position, and the ratio ${N_{\klgg}^{far}}/{N_{\klppp}^{far}}$ 
measured from the far-target beam allow
the direct evaluation of the number of $\klgg$ decays in the
near-target beam. By using the same ranges of 
energy and decay position only a
very small acceptance correction is needed due to the slight differences in
geometries of the two beams.
A by-product of this method is a precise measurement of $\ggtoppp$.

\section{\bf Experimental set-up and data taking}
 
The near-target beam contains $K^0$ mesons and hyperons.
It was produced on a 2 mm diameter, 400 mm long beryllium target by 400 GeV
protons at a production angle of 3.0 mrad in the vertical plane. 
In this special high intensity near-target mode, the primary proton beam was
attenuated and collimated to the desired intensity, far upstream of
the target.
The target is 
 followed by a sweeping magnet packed with tungsten-alloy inserts in which
protons not interacting in the target are absorbed, and by a 0.36 cm diameter
collimator. In the usual NA48 experimental configuration the 1.5 m long
collimator is followed successively by an anti counter (AKS), 
by an 89 m long evacuated
tube terminated by a thin Kevlar window 
and by a helium filled tank which contains the drift chambers and a
spectrometer magnet. 
However for these data the AKS counter, the Kevlar window and 
the drift chambers were removed and the helium tank was evacuated. 

The present analysis is based on data recorded 
during a 40-day run in 2000 with a
beam intensity of about $\sim 10^{10}$ protons hitting the target during the 
3.2 s long SPS spill.  This is a factor of $\sim300$ higher than the
usual near-target
beam. This near-target run was preceded by a 30-day run with
the far-target beam alone and the same detector
configuration\footnote{with exception of the proton beam energy which
  was 450 GeV, production angle which was 4.2 mrad 
 and of the spectrometer magnet which,
 unlike in the near-target run, 
 was powered in order to create occupancy
  conditions as close as possible to those of the direct CP-violation measurement.}, devoted mainly to studies 
of systematic effects in the direct CP-violation 
measurement~\cite{kmass,epspap}.
 
The detector elements used in the analysis are the following:
\bei
\item A liquid krypton calorimeter (LKr)~\cite{unal} is used to
  measure the energy, position and time of
electro-magnetic showers initiated by photons ($\gamma$). Around 20 t of liquid
krypton at 121 K are used as an ionization detector. 
The calorimeter has a structure of 13212 square towers of $2\times2$ cm$^2$
cross-section and 127 cm length (27 radiation lengths) each. The cells are 
formed by copper-beryllium ribbons, 1.8 cm  wide and 40 $\mu$m thick, stretched 
longitudinally. The ionization signal from each of the cells is
amplified, shaped, and digitised by 10-bit FADCs at 40 MHz sampling
frequency. The dynamic range of the digitisers is increased by
gain-switching amplifiers which change the amplification factor
depending on the pulse height.
The energy resolution is
\beq
\sigma(E)/E \simeq 0.090/E \oplus 0.032/\sqrt{E} \oplus 0.0042, \nonumber
\eeq
where E is in GeV. The read-out system was calibrated by a charge pulse every
burst during data taking. The final calibration of the energy scale is fixed by
fitting the effective edge of the collimator seen in the data to be
the same as in the Monte Carlo. The relative calibrations of the
individual cells were determined by using $K_{e3}$ decays during the 1998
run and checked to be similar in the 2000 run with $\pi^0,\eta
\rightarrow \gaga$ decays produced in thin targets in a special 
$\pi^-$-beam run. 
The position and time resolutions for a single photon with energy
larger than 20 GeV are better than 1.3 mm
and 300 ps, respectively.
\item A sampling hadron calorimeter composed of 48 steel plates, each 24 mm thick, interleaved with
scintillator planes is used to measure hadronic showers with a readout in horizontal and
vertical projections.
\item A scintillator hodoscope upstream of the two calorimeters helps
  to detect charged particles. It is composed of two planes segmented
  in horizontal and vertical strips arranged in four quadrants.
\eei
A more complete description of the apparatus can be found elsewhere~\cite{epspap}.

The trigger decision,
common to $\gamma\gamma$, $\pipin$ and $\pi^0\pipin$ decays, is
based on quantities which are derived from the projections 
of the energy deposited in the
electro-magnetic liquid krypton calorimeter~\cite{nutnim}. The trigger required that the total deposited 
energy $E_{tot}$ is larger than 50 GeV, the radius of the centre of
energy is smaller than 15 cm and the proper life time of the kaon
is less than 9 $\ks$ lifetimes downstream of the collimator. The trigger
efficiency, $(99.8 \pm 0.1)$\%, is measured using a minimum-bias
sample triggered by a scintillating fibre hodoscope placed 
in the liquid krypton calorimeter and is equal for  $\gamma\gamma$,
$\pipin$ and $\pi^0\pipin$ decays in both the near- and far-target runs. 

\section{\bf Event selection }

The energies and positions of electro-magnetic showers initiated by
photons, measured in 
the liquid krypton calorimeter, are used to 
calculate the kaon energy and decay vertex. To select $\kgg$ candidates,
all events with $\geq 2$ energy clusters are considered. Pairs of
clusters which are in time within 5 ns and with no other 
cluster with energy $> 1.5$ GeV closer in time than 3 ns with respect
to the event time  are selected. The event time is computed from the
times of the two most energetic cells of the selected clusters.
In addition, the $\gamma\gamma$ pair must pass the following cuts:
\begin{itemize}
   \item The energy of each cluster must be greater than 3 GeV
         and less than 100 GeV.
   \item The transverse distance between two clusters is required to 
         be greater than 10 cm.
   \item The total energy of the selected cluster pair is required to be 
         less than 170 GeV and to be greater than 70 GeV.
   \item The radius of the centre of energy, 
\beq       
R_C = \frac{\sqrt{(\sum_i E_i x_i)^2+(\sum_i E_i y_i)^2}}
            {\sum_i E_i}
\eeq
         is required to be less than 7 cm, where $E_{i}$,
$x_{i}$, $y_{i}$ are the $i-{th}$ cluster energy, 
$x$ and $y$ coordinates at LKr, respectively.
   \item The energy deposited in the hadron calorimeter must not
         exceed 3 GeV in a time window of $\pm15$ ns around the event time.
\end{itemize} 
In order to determine the $K_S$ and $\kl$ fluxes in the beam, the decays 
$\kspp$ and $\klppp$ have been used. 
Similar conditions as for $\gamma\gamma$ cluster pairs
are applied to groups of 4 or 6 in-time clusters for $2\pi^0$ and
$3\pi^0$ respectively. 

The decay vertex position $z_{vertex}$ of a kaon decaying into photons
or $\pi^0$'s is calculated using the kaon mass constraint
\beq       
z_{vertex}= z_{Lkr} - \frac{\sqrt{\sum_{i,j,i>j}{E_i}{E_j}
                   [(x_i - x_j)^2 + (y_i - y_j)^2]}}{m_K}
\eeq

where $z_{Lkr}$ is the $z$ coordinate of the LKr calorimeter with respect
to the end of the collimator. 

In $2\pi^0$ and $3\pi^0$ samples a
$\chi^2$ cut of 27 ($\sim{5\sigma}$) is applied to the invariant
masses of photon pairs calculated using $z_{vertex}$.
For $2\pi^0$ decay, the $\chi^2$ variable is defined as
\beq
 \chi^2=\left[ \frac{(m_1+m_2)/2-m_{\pi^0}}{\sigma_+}\right]^2+
        \left[ \frac{(m_1-m_2)/2}{\sigma_-}\right]^2
\eeq
where $\sigma_{\pm}$ are the resolutions on $(m_1 \pm m_2)/2$ measured
from the data and parametrised as a function of the lowest photon energy.
For $\pipipin$ decays an equivalent variable is defined in the space of three
invariant masses.

Due to the large
background from $\kspp$ decays the study of $\ksgg$ can only
be carried out in a restricted decay region. The maximum
$\gamma\gamma$ mass from a $\kspp$ decay where two photons escape
detection is 458 MeV/c$^2$ which would
lead to an apparent vertex shift of 9 m downstream. 
However, in case some showers overlap and only one or no photon
misses the detector this shift can be significantly smaller. Therefore
the decay region is
defined to be between -1 m and 5 m with respect to the exit of the
collimator.

In order to remove events with hadronic or 
overlapping showers from the $\gamma\gamma$ sample, 
the shower width is required to be less than
3$\sigma$ above the average value for photon showers of a given energy. 
This cut, which
is calibrated from showers in $\kspp$ decays, 
removes $<0.5\%$ of good $\kgg$ events.

In the near-target data, 19819 events passed the $\gaga$ selection cuts
while in the far-target data more than $6\times 10^4$ events were
used. Normalisation channels were down-scaled, retaining 
several hundred thousand events for analysis.

\section{Background subtraction}

The $\pipin$ and $\pipipin$ samples are essentially background free. In the
$\gamma\gamma$ samples the following backgrounds have been identified and
subtracted:
\begin{itemize}
\item A small hadronic background visible as a tail in the shower
  width distribution. In the 
  near-target sample this background originates 
 from interactions of the beam particles in
  the collimator. In the far-target beam, which is cleaned by
  three collimators, the background from this source is negligible. 
  However, pairs
  of charged hadrons generated in the far-target beam upstream of
  or inside the
  final collimator and
  opened vertically by the last sweeping magnet have been
  found to contribute to the $\klgg$ sample.
\item Accidental $\gamma\gamma$ pairs visible as a flat continuum 
  in the distribution of the shower time differences. 
\item Residual $\kspp$ decays with one shower overlap and one
  low-energy photon missing the calorimeter.
\item Dalitz decay $K^0\rightarrow ee\gamma$ with an overlapping $ee$
  pair or one shower with energy below the detection limit of 1.5 GeV.
  In the far-target run, due to the presence of
      magnetic field, $ee$ pairs do not overlap.
\end{itemize}
The last three background sources contribute significantly only to the
near-target $\gaga$ sample.

The hadronic and accidental backgrounds are subtracted using the 
distribution in radius of the centre
of energy ($R_C$), which indicates the deviation of the direction
of the apparent decay source from the beam axis. A control region
is defined with $R_C$ between 8 cm and 10 cm. 
In order to obtain the background-extrapolation
factor to the signal region, $R_C<7$ cm, a set of model distributions
has been extracted from the data by modifying certain selection requirements.

\begin{figure}[ht]
\begin{center}
\mbox{\epsfig{file=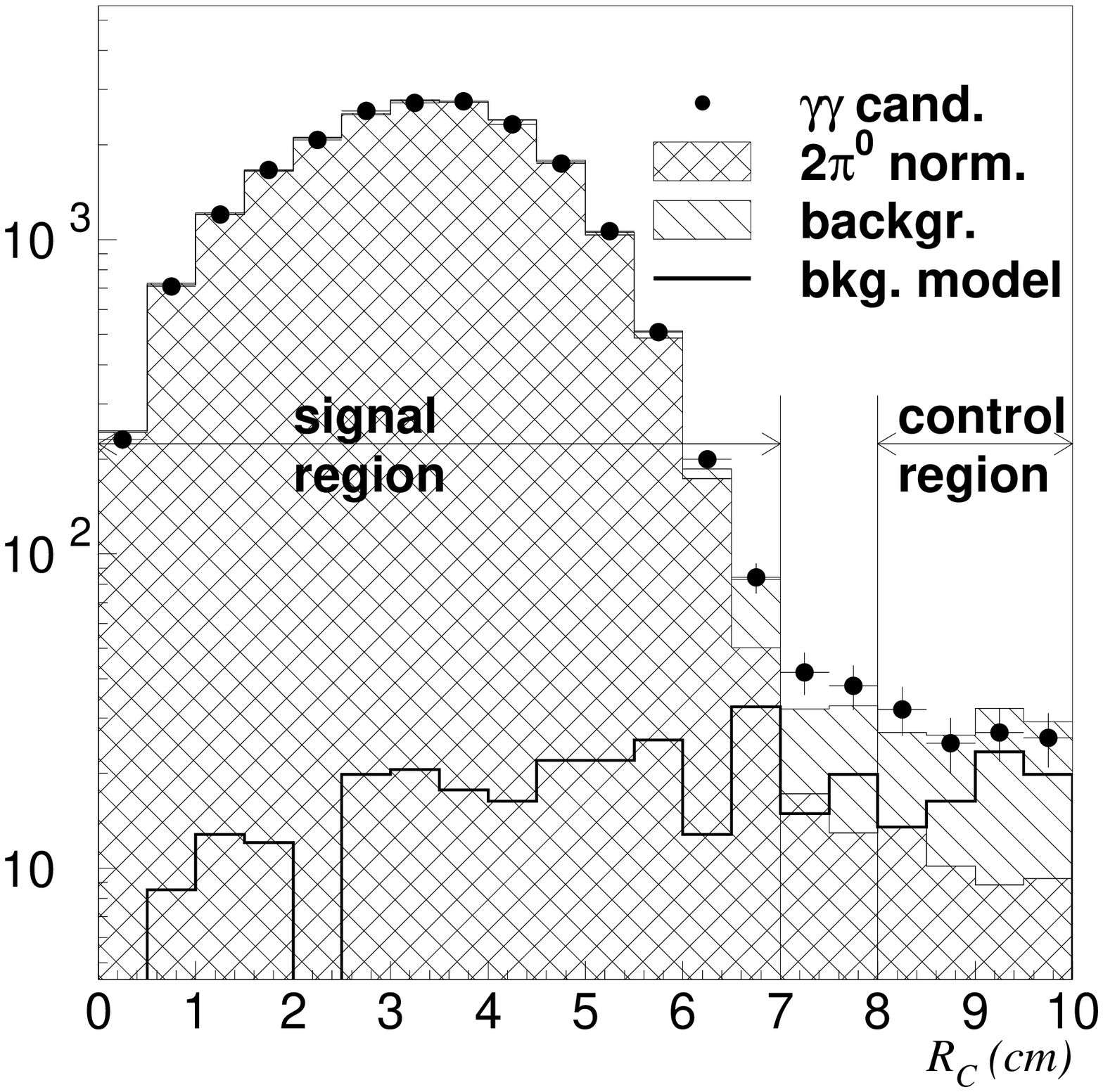,width=6cm}}
\mbox{\epsfig{file=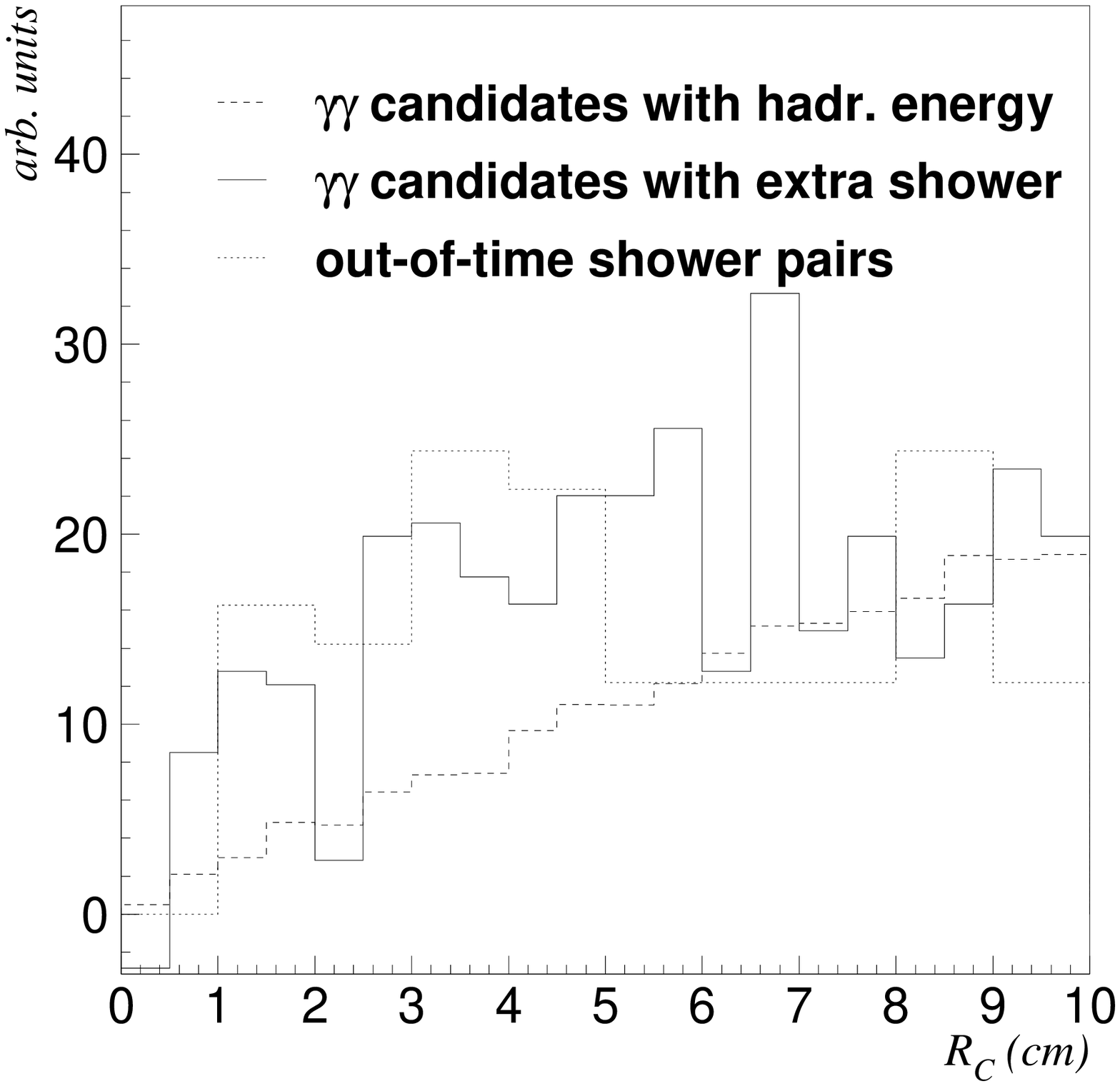,width=6cm}}
\end{center}
\caption{ {\em Left:} Example of background
  subtraction in the near-target $\gaga$ sample using as a model
  $\gaga$ candidates with an extra shower.
{\em Right:} Comparison of the centre-of-energy distributions 
of background models 
  in the near-target data. }
\label{fig:bkg}
\end{figure}

In the near-target data, in order to study the hadronic background
component two different samples have been selected. 
In one, a hit in the hadron calorimeter is required and the shower 
width cut is released (extrapolation factor 1.5). In the other sample,
as collimator scattering can produce more than two particles, 
an additional shower in the calorimeter is required and the 
decay region is shortened by 1 m in order to avoid contamination from
$\kspp$ decays, while all other cuts are unchanged 
(extrapolation factor 3.1) (Fig.~\ref{fig:bkg} left).
The model of accidental background is obtained by requiring
the two showers to be out of time by more than
  5 ns and less than 10 ns (extrapolation factor 2.7).
The median value of the extrapolation factors from three different
models gives
an inclusive hadronic and accidental background estimate of
$(2.1 \pm 0.7)$\% of the $\ksgg$ sample 
where the uncertainty corresponds to the spread of
the extrapolation factors (Fig.~\ref{fig:bkg} right).

\begin{figure}[ht]
\begin{center}
\mbox{\epsfig{file=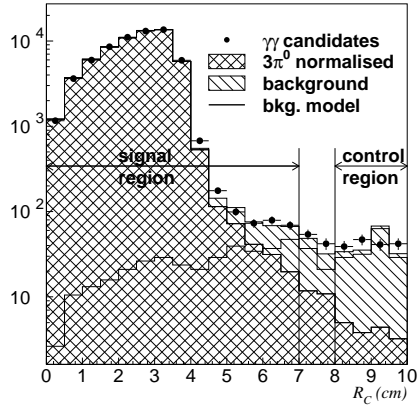,width=6cm}}
\end{center}
\caption{ Background
  subtraction in the far-target $\gaga$ sample.}
\label{fig:bkgl}
\end{figure}

In the far-target data the background model sample is obtained by
requiring
a hit in the hadron calorimeter and a hit in the
  scintillator hodoscope giving an extrapolation factor
  of 2.3. Using this factor and a conservative uncertainty leads
  to a background estimate of $(0.6 \pm 0.3)$\% (Fig.~\ref{fig:bkgl}). In both
  background estimates the amount of beam halo
  in the control region is taken into account and
  is estimated from $\pipin$ and
  $\pipipin$ samples.

\begin{figure}[ht]
\begin{center}
\mbox{\epsfig{file=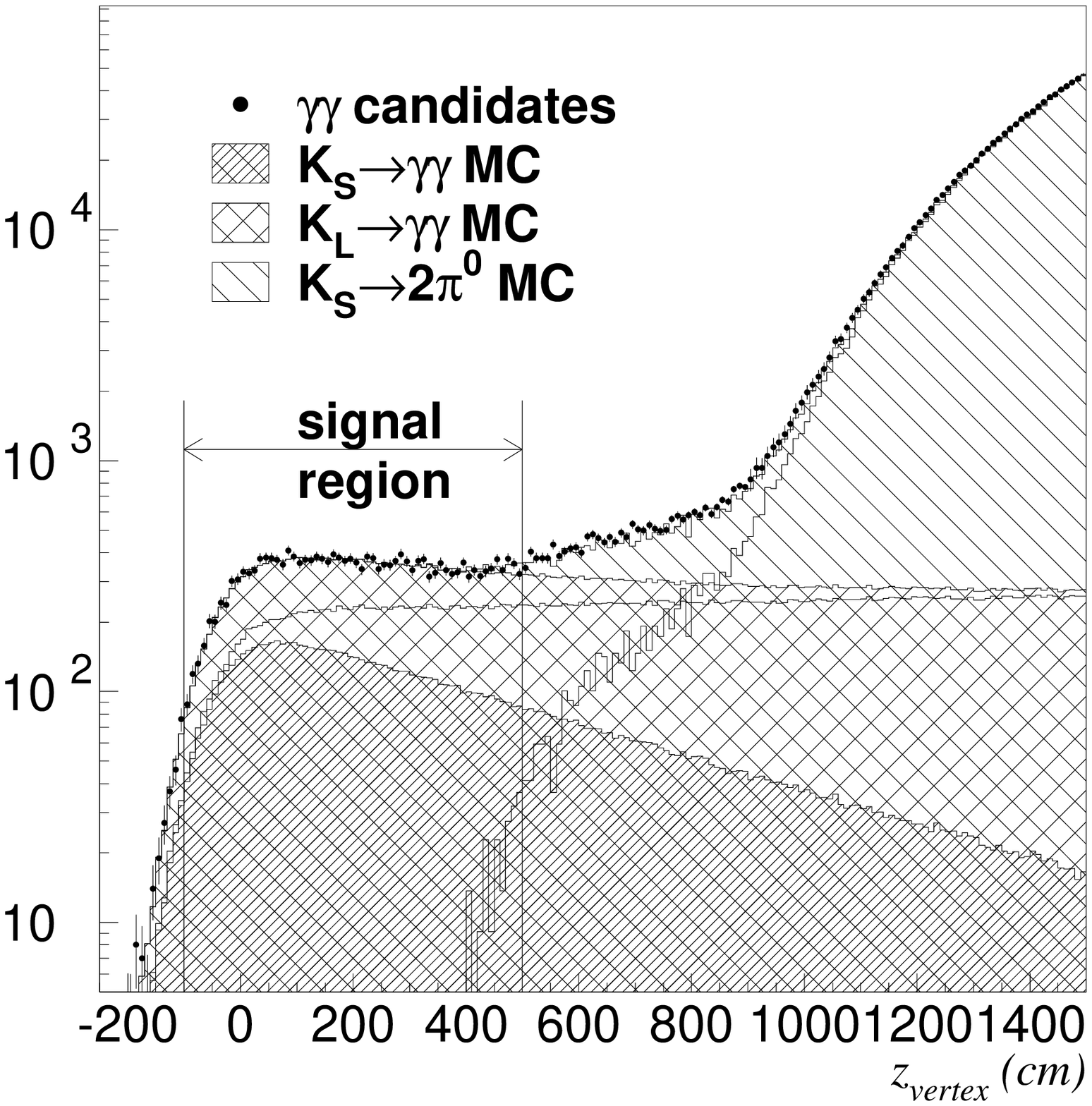,width=6.5cm}}
\mbox{\epsfig{file=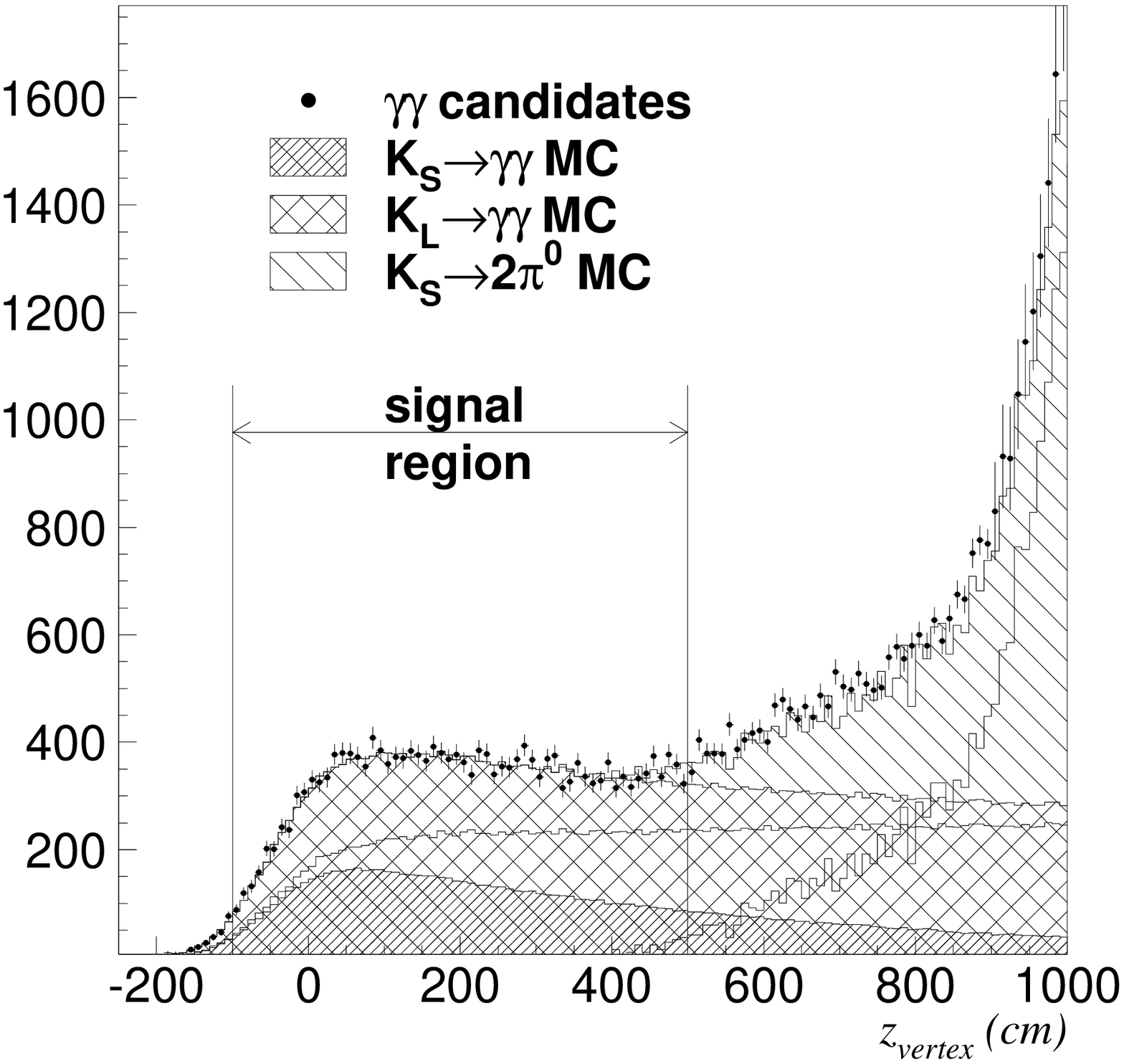,width=6.5cm}}
\end{center}
\caption{The $z_{vertex}$ distribution of the near-target $\gamma\gamma$ sample
  with logarithmic (left) and linear (right) scales
  compared to Monte Carlo samples of $\ksgg$, $\klgg$ and $\kspp$ background,
 normalised to the flux and using the measured values of 
$\ggtoppp$ and BR($\ksgg$). The shapes of the distributions before 
  summing are also indicated.}
\label{fig:zplot}
\end{figure}

In order to subtract the $\kspp$ background a sample of 
$1.2 \times 10^9$ decays
was generated by Monte Carlo. The Monte Carlo sample was normalised using
fully reconstructed  $\kspp$ events and checked by comparing it to the
data in the 10~m following the decay region where $\kspp$ background
largely dominates the $\gamma\gamma$ sample (Fig.~\ref{fig:zplot}). 
The estimated background
amounts to $(2.1 \pm 0.4)$\% of the $\ksgg$ sample
and populates mainly the last
metre of the decay region. The uncertainty reflects a
discrepancy between data and Monte Carlo in the simulation of
overlapping showers.

The contribution of $K^0\rightarrow ee\gamma$ decays to the
$\gamma\gamma$ sample was estimated using Monte Carlo to be 0.4\%. 
In the same way,
the numbers of accepted $\pipin$ (0.9\%) and $\pipipin$ (1.1\%) decays with a
subsequent $\pi^0 \rightarrow ee\gamma$ decay with respect to pure
photon final states were estimated. 
The overall correction to the $BR(\ksgg)$ due to accepted Dalitz decays
amounts to $(1.5 \pm 0.3)$\%, where the uncertainty is due to the 
simulation of overlapping showers.

\section{\bf Results}

Binned maximum likelihood fits have been used to measure 
the $\ggtoppp$ ratio and the $\ksgg$ branching ratio 
in 10 kaon energy and 6 longitudinal vertex position bins.
Acceptances were
calculated using Monte Carlo with detector simulation based on
GEANT~\cite{geant} where the LKr calorimeter response was simulated
using a pregenerated shower library.  

The ratio of $\klgg$ and $\klppp$ decay rates, calculated from
far-target samples and corrected for acceptance and
for hadronic background as described in the
previous section, is
\beq
\label{eq:ggtoppp}
\ggtoppp = (2.81 \pm 0.01_{stat} \pm 0.02_{syst}) \times 10^{-3}
\eeq 
where the systematic uncertainty accounts for the uncertainties in the
background subtraction and simulation of the calorimeter response in
the acceptance calculation. The latter were checked 
by varying selection cuts,
especially the shower-width cut in the $\gamma\gamma$ sample and
the $\chi^2$ cut in the $\pipipin$ sample, 
which are specific to each of these
decay channels. In addition it was checked that the result is insensitive
to various parameters, e.g. longitudinal vertex position and kaon energy.

The number of $\klgg$ events in the near-target $\gaga$ sample
is given by
\beq
  N_{\gaga}^{near} = R_{A} \frac{N_{\klgg}^{far}}{N_{\klppp}^{far}} 
                 N_{\klppp}^{near}
\eeq

where $N$ are the background corrected numbers of events from the 
near and far-target runs. 
$R_A$ is a double ratio of acceptances
\beq
  R_A = \frac{A_{\klgg}^{near}}{A_{\klppp}^{near}} /
        \frac{A_{\klgg}^{far}}{A_{\klppp}^{far}}
\eeq
which is close to unity ($\sim$1.05) because the same decay region was
used for the
far and near-target samples. For this reason only the background
subtraction uncertainty in (\ref{eq:ggtoppp}) enters the calculation
of BR($\ksgg$), while the acceptance calculation uncertainty cancels
between near and far-target samples.

The summary of all corrections and uncertainties 
in the BR($\ksgg$) calculation is given in
Table~\ref{tab:sys}. The acceptance uncertainty 
reflects the level of knowledge of non-Gaussian
tails in the energy reconstruction accounted for in the acceptance
calculation. 

\begin{table}
\begin{center}
\begin{tabular}{|l|r|}
\hline 
Hadronic background and accidentals & -2.1 $\pm$ 0.7 \% \\
Background from $\ks \rightarrow \pipin$ & -2.1 $\pm$ 0.4 \% \\
Had. background in far-target beam & 0.9 $\pm$ 0.4 \% \\
Dalitz decay correction & 1.5 $\pm$ 0.3 \% \\
Acceptance uncertainty systematic & $\pm$ 0.3 \% \\
Acceptance uncertainty statistical & $\pm$ 0.6 \% \\
Trigger efficiency & $\pm$ 0.1 \% \\ \hline
Total & { -1.8 $\pm$ 1.2 \%} \\ \hline
Statistical uncertainty data & { $\pm$ 2.0 \%} \\
BR($\ks \rightarrow \pipin$) & $\pm$ 0.9 \% \\ \hline
\end{tabular}
\caption{Summary of all corrections and uncertainties in
  the BR($\ksgg$) measurement.}
\label{tab:sys}
\end{center}
\end{table}

From 19819 $\gaga$ candidates, $7461 \pm 172$ have been estimated to be $\ksgg$
decays, giving a ratio of decay rates
\beq
  \frac{\Gamma(\ksgg)}{\Gamma(\kspp)} = 
            (8.84 \pm 0.18_{stat} \pm 0.10_{syst})
         \times 10^{-6}
\eeq
Using BR($\ks \rightarrow \pipin$) taken from \cite{pdg} one obtains
\begin{eqnarray}
  BR(\ksgg) &=& (2.78 \pm 0.06_{stat} \pm 0.03_{syst} \pm 0.02_{ext})
         \times 10^{-6} \\
            &=& (2.78 \pm 0.07) \times 10^{-6}
\end{eqnarray}
where the external uncertainty is that quoted in Table~\ref{tab:sys}
for the $\kspp$ branching ratio.
Systematic checks have confirmed that the result is
stable with respect to longitudinal vertex position, kaon energy, run
period and other parameters.

This result is in agreement with previous measurements,
but has much better precision. 
It is the first measurement of $BR(\ksgg)$ 
 to show significant difference with respect to the
$O(p^4)$ calculations of ${\chi}PT$ and indicates that
higher order corrections increase the decay rate by about 30\%.

\section*{ACKNOWLEDGEMENT}
It is a pleasure to thank the technical staff of the participating
laboratories, universities and affiliated computing centres for 
their efforts in the construction of the NA48 apparatus, in the 
operation of the
experiment, and in the processing of data. We would also like to thank
G.~D'Ambrosio for useful discussions concerning ${\chi}PT$ and for
clarifying the precision of the theoretical prediction for the $\ksgg$
decay.

\end{document}